\documentclass[3p,times]{elsarticle}
\pdfoutput=1
\usepackage{ecrc}

\volume{00}

\firstpage{1}

\runauth{}

\usepackage{amsmath}
\usepackage{latexsym}
\usepackage{amsfonts}
\usepackage{graphicx}
\usepackage{mathrsfs}
\usepackage{epstopdf}
\usepackage{orcidlink}
\usepackage{comment}
\begin{document}

\begin{frontmatter}
\title{Approximate reconstruction of inflationary potential with ACT observations}

\author[labela]{Zhu Yi \orcidlink{0000-0001-7770-9542}}
\address[labela]{Faculty of Arts and Sciences, Beijing Normal University, Zhuhai 519087, China}
\ead{yz@bnu.edu.cn}

\author[labela]{Xingzhi Wang}
\ead{wangxingzhi@mail.bnu.edu.cn}

\author[labelb]{Qing Gao \corref{cor1} \orcidlink{0000-0003-3797-4370}}
\address[labelb]{School of Physical Science and Technology, Southwest University, Chongqing 400715, China}
\ead{gaoqing1024@swu.edu.cn}
\cortext[cor1]{Corresponding author}

\author[labelc]{Yungui Gong \orcidlink{0000-0001-5065-2259}}

\address[labelc]{Institute of Fundamental Physics and Quantum Technology, Department of Physics, School of Physical Science and Technology, Ningbo University, Ningbo, Zhejiang 315211, China}
\ead{gongyungui@nbu.edu.cn}

\begin{abstract}
The Atacama Cosmology Telescope (ACT) has recently reported updated measurements of the scalar spectral index $n_s$,
revealing a tension with the predictions of many conventional inflationary models. 
In this work, we adopt a parameterization of the spectral index in the form $n_s = 1 - p/(N + \alpha)$ with $1.338<p<1.746$,
to reconstruct an inflationary potential consistent with the latest ACT data.
The resulting potential is  the  Kachru–Kallosh–Linde–Trivedi (KKLT) potential $V(\phi) = V_0/[1 + (M/\phi)^n]$,
where the power index is given by $n = 2(p-1)/(2-p)$.
The corresponding tensor-to-scalar ratio is approximately
$r \approx  16(p-1)/[C (N+\alpha)^p]$ with $C= 2^{2p-1} \left[\sqrt{p-1}/(2-p)\right]^{2p-2} /M^{2p-2}$.
Since the reconstruction under the slow-roll approximation is model-independent, the KKLT model can serve as an effective approximation to a broad class of inflationary scenarios that are consistent with the latest ACT measurements of  $n_s$, at least on large scales.

\end{abstract}

	
\end{frontmatter}
\section{introduction}
Inflation provides a compelling resolution to several long-standing issues in the standard big bang cosmology, including the flatness, horizon, and monopole problems, while simultaneously generating the primordial density fluctuations that seed the formation of large-scale structures \cite{Starobinsky:1980te, Guth:1980zm, Linde:1981mu, Albrecht:1982wi}. Quantum fluctuations of the inflation field give rise to metric perturbations, which are imprinted as temperature anisotropies in the cosmic microwave background (CMB). Two key observables characterizing these perturbations are the scalar spectral index $n_s$, which quantifies the scale dependence of scalar modes, and the tensor-to-scalar ratio $r$, which measures the amplitude of tensor modes relative to scalar ones.

Recent observations from the Atacama Cosmology Telescope (ACT),  when combined with other observational data  \cite{ACT:2025fju, ACT:2025tim}, report a higher value of $n_s$ compared to earlier results from the {\it Planck} satellite \cite{Planck:2018jri}. 
A joint analysis combining ACT and {\it Planck} data (P-ACT) yields $n_s = 0.9709 \pm 0.0038$, 
while the inclusion of CMB lensing and DESI BAO measurements (P-ACT-LB) further raises this to $n_s = 0.9743 \pm 0.0034$ \cite{DESI:2024uvr, DESI:2024mwx}. 
This represents a $\sim 1\sigma$ upward shift relative to the {\it Planck}-only constraint and has significant implications for inflationary model building. 
The upward shift in the scalar spectral index $n_s$ resulting from the addition of DESI BAO data may arise from a tension between DESI BAO and CMB data in the measurements of $\Omega_{m}$ and $r_h h$ using the $\Lambda$CDM model, 
since $n_s$ is degenerate with $\Omega_{m}$ and $r_h h$ \cite{Ferreira:2025lrd}.

A wide class of inflationary scenarios, including hilltop models \cite{Boubekeur:2005zm}, $T$ and $E$ models \cite{Kallosh:2013hoa, Kallosh:2013maa}, $R^2$ inflation \cite{Starobinsky:1980te}, and Higgs inflation with strong nonminimal coupling \cite{Kaiser:1994vs, Bezrukov:2007ep}, predict a universal attractor behavior for the scalar spectral index: $n_s = 1 - 2/N$, where $N$ denotes the number of $e$-folds between horizon exit and the end of inflation. For $N = 60$, this yields $n_s \approx 0.9667$, which closely matches the {\it Planck} central value but falls below the P-ACT-LB $1\sigma$ constraints. Consequently, these attractor models are now under large observational tension, motivating the exploration of alternative scenarios that can accommodate the higher value of $n_s$ indicated by the latest data.

Various approaches have been explored to alleviate this tension, such as relaxing the strong coupling limit in the inflation with nonminimal coupling $\xi f(\phi)R$ \cite{Kallosh:2025rni, Gao:2025onc}, accounting for reheating effects \cite{Liu:2025qca, Haque:2025uis, Zharov:2025evb, Haque:2025uri, Drees:2025ngb,Ballardini:2024ado}, and considering other theoretical frameworks \cite{He:2025bli, Gialamas:2025kef, Frob:2025sfq, Dioguardi:2025vci, Brahma:2025dio, Berera:2025vsu, Aoki:2025wld, Dioguardi:2025mpp, Salvio:2025izr, Yogesh:2025wak, Gialamas:2025ofz,Gao:2025viy,Peng:2025bws,Odintsov:2025jky,Oikonomou:2025htz,Odintsov:2025eiv,Zhu:2025twm}.  

In this work, we adopt another approach based on inflationary potential reconstruction within the single-field slow-roll framework.
Unlike traditional methods,  which begin with 
a specific inflationary model and derive its predictions,  the reconstruction approach begins with a parametrization of observable quantities as functions of the $e$-folding number $N$, typically $n_s(N)$ and $r(N)$, and employs the relations between these observables and the background/perturbation dynamics to reconstruct 
the corresponding inflationary potential $V(\phi)$ \cite{Mukhanov:2013tua, Roest:2013fha, Garcia-Bellido:2014gna, Creminelli:2014nqa, Barranco:2014ira, Galante:2014ifa, Chiba:2015zpa, Lin:2015fqa, Yi:2016jqr, Odintsov:2016vzz, Odintsov:2017qpp, Nojiri:2017qvx, Gao:2017uja, Gao:2017owg, Fei:2017fub, Fei:2020jab}. For example, the universal attractor form $n_s = 1 - 2/N$ leads directly to the $T$-model potential under the reconstructed method \cite{Lin:2015fqa}. 

This method is nearly model-independent, as it makes minimal assumptions about the specific functional form of the potential $V(\phi)$, requiring only that the potential be sufficiently flat. The reconstructed potential is fully determined by the chosen functional form of the observational parameters, such as $n_s(N)$ or $r(N)$.
To ensure consistency with observational data, we consider a generalized form of the scalar spectral index, $n_s(N) = 1 - {p}/{(N + \alpha)}$, where $\alpha\ll N$ and the parameter $p$ quantifies deviations from the universal attractor behavior. This form provides sufficient flexibility to reconstruct inflationary potentials consistent with the latest P-ACT-LB constraints.

The remainder of this paper is organized as follows. In Sec.~\ref{sec:rec}, we briefly review the reconstruction framework. In Sec.~\ref{sec:pot}, we apply this method to derive inflationary potentials consistent with the latest data. Our conclusions are summarized in Sec.~\ref{sec:con}.


\section{The reconstruction method}
\label{sec:rec}
For a canonical single-field inflation model with potential $V(\phi)$, the slow-roll parameters are defined by
\begin{equation}
    \epsilon_V \equiv \frac{1}{2}\left(\frac{V'}{V}\right)^2, \quad \eta_V \equiv \frac{V''}{V},
\end{equation}
where primes denote derivatives with respect to the inflation field $\phi$, i.e., $V' = dV/d\phi$ and $V'' = d^2V/d\phi^2$. Throughout this work, we set the reduced Planck mass to unity, $M_{\mathrm{pl}} = 1/\sqrt{8\pi G} = 1$.

Under the slow-roll approximation, $\epsilon_V \ll 1$ and $|\eta_V| \ll 1$, the scalar spectral index $n_s$ and tensor-to-scalar ratio $r$ are given at leading order by
\begin{equation}
\label{nsapproxeq2}
n_s - 1 \approx 2\eta_V - 6\epsilon_V, \quad r \approx 16\epsilon_V.
\end{equation}
An auxiliary relation between the slow-roll parameters is 
\begin{equation}
\label{nsapproxeq3}
\frac{d\ln\epsilon_V}{dN} = 2\eta_V - 4\epsilon_V,
\end{equation}
which allows Eq.~\eqref{nsapproxeq2} to be rewritten as
\begin{equation}
\label{nsapproxeq4}
n_s - 1 = -2\epsilon_V + \frac{d\ln \epsilon_V}{dN},
\end{equation}
where $N$ denotes the number of $e$-folds remaining before the end of inflation when the pivotal scale exits the horizon.

To relate the potential to $N$, we begin with the energy conservation equation for the inflation,
\begin{equation}
\label{sclreq1}
d\ln\rho_\phi + 3(1 + w_\phi) d\ln a = 0,
\end{equation}
where $\rho_\phi = V(\phi) + \dot{\phi}^2/2$ is the inflation energy density and $w_\phi = (\dot{\phi}^2/2 - V)/(\dot{\phi}^2/2 + V)$ is the equation-of-state parameter. Under the slow-roll condition,  the kinetic term is subdominant, so that $\rho_\phi \approx V(\phi)$. The  equation-of-state parameter can be expressed in terms of the slow-roll parameter, 
\begin{equation}
\label{sclreq6}
1+w_\phi= \frac{\dot\phi^2}{V(\phi)}\approx \frac{2}{3}\epsilon_V.
\end{equation}
Combining Eqs.  \eqref{sclreq1}  and \eqref{sclreq6}, and using relation $d\ln a =-d N$, we have 
\begin{equation}
\label{sclreq2}
\epsilon_V \approx \frac{1}{2} \frac{d\ln V}{dN} > 0.
\end{equation}

By substituting Eq.~\eqref{sclreq2} into Eq.~\eqref{nsapproxeq4}, we obtain an expression for $n_s$ in terms of the potential and $N$  \cite{Chiba:2015zpa}:
\begin{equation}
\label{nsapproxeq6}
n_s - 1 \approx -(\ln V)_{,N} + \left(\ln \frac{V_{,N}}{V} \right)_{,N} = \left(\ln \frac{V_{,N}}{V^2} \right)_{,N}.
\end{equation}
These relations indicate that the functions $n_s(N)$, $\epsilon_V(N)$, and $V(N)$ are related,  so knowing any one of them, we can  determine the others.

To reconstruct the inflationary potential as a function of the field, $V(\phi)$, one needs to relate the inflation field   $\phi$ to the number of $e$-folds $N$. Using Eq.~\eqref{sclreq6} along with $H^2 \approx V/3$, the field evolution equation becomes 
\begin{equation}
\label{nphieq1}
\frac{d\phi}{dN} \approx \pm \sqrt{2\epsilon_V(N)}, 
\end{equation}
where the sign depends on the direction of field evolution.  Integrating Eq.~\eqref{nphieq1}, we obtain
\begin{equation}
\label{nphieq2}
\phi - \phi_e \approx  \pm \int_0^N \sqrt{2\epsilon_V(N)} \, dN,
\end{equation}
with $\phi_e$ denoting the field value at the end of inflation, i.e., $\phi_e=\phi(N=0)$. The integral may be evaluated analytically or numerically, depending on the form of $\epsilon_V(N)$. 

Note that the above reconstruction process is based on slow-roll approximation,  
so the relations \eqref{nsapproxeq6}-\eqref{nphieq2} only provide a reliable approximation in the slow-roll regime $\epsilon_V \ll 1 $, and discrepancies may arise near the end of inflation where $ \epsilon_V \sim 1 $. 
Since we mainly focus on the behavior of the inflationary potential on CMB scales, 
the number of $e$-folds $N\sim 50-60$, so the slow-roll approximation is well satisfied.
Therefore, the reconstruction method is valid in the large $N$ limit,
and the reconstructed potential approximates a broad class of inflationary potentials consistent with the CMB observations only at CMB scales.

In standard approaches,  one begins with a specific form of $V(\phi)$ and derives predictions for the inflationary observables. In contrast, the reconstruction methods allow one to start from observable quantities, such as $n_s(N)$ or $r(N)$, and reconstruct the inflationary potential $V(\phi)$.   Given any one of the functions $n_s(N)$, $\phi(N)$, $r(N)$, or $V(N)$, the remaining quantities can be consistently determined.

\section{The Potential}
\label{sec:pot}

A joint analysis of the {\it Planck}, DESI, and ACT data provides the following constraint on the scalar spectral index \cite{ACT:2025fju}:
\begin{equation}\label{ns:constraint}
    n_s = 0.9743 \pm 0.0034.
\end{equation}
These observations disfavor the so-called universal attractor scenario, where the scalar spectral index has the form
\begin{equation}
    n_s - 1 = -\frac{2}{N},
\end{equation}
for a typical number of $e$-folds $N = 60$. This universal attractor arises in a variety of inflationary models, such as Higgs inflation with nonminimal coupling, $R^2$ inflation, hilltop inflation, and E/T models. Consequently, these models are   incompatible with current observational constraints, motivating the need to identify a potential consistent with data.

To reconstruct a compatible potential, we adopt a parametric form for the scalar spectral index:
\begin{equation}\label{ns:para}
    n_s = 1 - \frac{p}{N + \alpha},
\end{equation}
where $\alpha$ is small and introduced to regularize the expression near $N = 0$.   To match the constraint \eqref{ns:constraint}, the parameter $p$ must satisfy
\begin{equation}\label{p:constraint}
    1.338 < p < 1.746,
\end{equation}
assuming $N = 60$ and $\alpha \approx 0$.

Combining the parametrization \eqref{ns:para} with the slow-roll expression \eqref{nsapproxeq4}, we obtain \cite{Lin:2015fqa}
\begin{equation}
\label{epsilonneq1}
\epsilon_V(N) = \frac{p - 1}{2(N + \alpha) + C(N + \alpha)^p},
\end{equation}
where $C \geq 0$ is an integration constant. 
At the end of inflation, we have $\epsilon_V(N=0) \approx 1$, the integration constant $C$ can be expressed as 
\begin{equation}
    \label{paramrel1}
    C \approx \frac{p - 1 - 2\alpha}{\alpha^p}.
\end{equation}
From Eqs.  \eqref{epsilonneq1} and \eqref{nsapproxeq2}, the tensor-to-scalar ratio is
\begin{equation}\label{res:r}
    r = \frac{16(p - 1)}{2(N + \alpha) + C(N + \alpha)^p}.
\end{equation}
If $C = 0$, then under the constraint \eqref{p:constraint} and for $N = 60$, we find $r \approx 8(p - 1)/(N + \alpha) > 0.045$, which exceeds the observational upper bound $r < 0.036$ \cite{BICEP:2021xfz}. Hence, we restrict our analysis to the  case $C > 0$.

From Eq. \eqref{ns:para}, the running of the spectral index is
\begin{equation}\label{rns}
    n_s' =-\frac{d n_s}{d N}=- \frac{p}{(N + \alpha)^2},
\end{equation}
so $n_s'<0$. For $N = 60$ and $\alpha \approx 0$, 
using the constraint \eqref{p:constraint}, 
we obtain
$-4.85\times 10^{-4} < n_s' < -3.72\times 10^{-4}$.
Although the result is in a little tension with the observation \cite{ACT:2025tim}
\begin{equation}
\label{rnsr}
    n_s' =0.0062\pm 0.0052, 
\end{equation}
at the $1\sigma$ level, it is still consistent with the observation at the $2\sigma$ level.

Substituting the parameterization  \eqref{ns:para} into the relation \eqref{nsapproxeq6} and solving it, we reconstruct the potential \cite{Chiba:2015zpa,Lin:2015fqa} as  
\begin{equation}
\label{vpareq1}
V(N) = \frac{p - 1}{A} \left[ \frac{1}{(N + \alpha)^{p - 1}} + \frac{C}{2} \right]^{-1},
\end{equation}
with integration constant $A > 0$. To derive the potential $V(\phi)$, we relate $\phi$ and $N$. Substituting Eq. \eqref{epsilonneq1} into the
relation \eqref{nphieq2}, we have  
\begin{equation}
\label{phineq6}
\begin{split}
\phi(N) &= \phi_0 \pm \frac{2}{2 - p} \sqrt{ \frac{2(p - 1)}{C} } (N + \alpha)^{1 - p/2} \times \\
&\quad {}_2F_1\left[ \frac{1}{2}, \frac{p - 2}{2(p - 1)}, \frac{4 - 3p}{2 - 2p}, -\frac{2}{C(N + \alpha)^{p - 1}} \right],
\end{split}
\end{equation}
where $_2F_1$ is the hypergeometric function and $\phi_0$ is an integration constant, which can relate to $\phi_e$ from the condition $\phi_e=\phi(N=0)$. 
For $|z| < 1$, the hypergeometeric function can be expressed as the power series
\begin{equation}
    _2F_1(a,b;c;z) = \sum_{k=0}^\infty \frac{(a)_k(b)_k}{(c)_k} \frac{z^k}{k!},
\end{equation}
where the Pochhammer symbol $(a)_k$ is defined as
\begin{equation}
  (a)_k=\begin{cases}
  1, & k=0, \\
  a(a+1) \cdots(a+k-1), & k>0. 
  \end{cases}
\end{equation}
 
To obtain a simplified expression of the relation $\phi$ and $N$, we consider  $C > 1$,  which is achievable for sufficiently small $\alpha$ according to Eq.~\eqref{paramrel1}. 
In this case, under the constraint \eqref{p:constraint}, Eq.~\eqref{phineq6} simplifies to
\begin{equation}\label{phiN:simple}
\phi(N) \approx \phi_0 \pm \frac{2}{2 - p} \sqrt{ \frac{2(p - 1)}{C} } (N + \alpha)^{1 - p/2}.
\end{equation}
Substituting it into \eqref{vpareq1} and assuming  $\phi_0=0$ and $\phi > 0$, the potential becomes
\begin{equation}
\label{brane1}
    V(\phi) = \frac{V_0}{1 + (M/\phi)^n},
\end{equation}
where  the power index $n$ is 
\begin{equation}\label{index:np}
n=2(p-1)/(2-p),
\end{equation}
and  the  parameters $V_0$, $M$ are given by
\begin{equation}\label{VM:parameter}
V_0 = \frac{2(p-1)}{A C}, \quad M =\frac{2^{\frac{2p-1}{2p-2}}  \sqrt{p-1} }{(2-p)C^{\frac{1}{2p-2}}}.
\end{equation}
The potential \eqref{brane1} is also known as  Kachru–Kallosh–Linde–Trivedi (KKLT) potential \cite{Kachru:2003aw,Martin:2013tda}.
Combining Eqs. \eqref{VM:parameter} and \eqref{phiN:simple}, we find 
\begin{equation} \label{eq:mphi}
    \left( \frac{M}{\phi} \right)^n = \frac{2}{C(N + \alpha)^{p - 1}} \ll 1,
\end{equation}
so the reconstructed potential \eqref{brane1} can be approximated as the inverse-power plateau potential \cite{Garcia-Bellido:2014gna,Creminelli:2014nqa, Dioguardi:2023jwa}
\begin{equation}
\label{rec:potential}
    V(\phi) = V_0 \left[ 1 - \left( \frac{M}{\phi} \right)^n \right].
\end{equation}
The model is also known as D-brane inflation or polynomial $\alpha$-attractors \cite{Dvali:2001fw,Martin:2013tda,Kallosh:2019jnl,Kallosh:2022feu}.
Note that the potential \eqref{rec:potential} is singular at $\phi=0$ and unbounded from below for $n>0$. 
However, the consistent generalization of the potential is the KKLT potential \eqref{brane1} \cite{Kachru:2003aw,Kallosh:2019jnl} we reconstructed. 
We will focus on the reconstructed KKLT potential \eqref{brane1}.
At the leading order, from  the parameterization \eqref{ns:para} and the result \eqref{res:r}, 
the KKLT potential \eqref{brane1} yields the following predictions,
\begin{equation}\label{nsr:para}
    n_s \approx 1 - \frac{p}{N}, \quad
    r \approx \frac{16(p - 1)}{C N^p},
\end{equation}
where the constant $C$ is
\begin{equation}
    C = 2^{2p - 1} \left[ \frac{\sqrt{p - 1}}{2 - p} \right]^{2p - 2} \frac{1}{M^{2p - 2}}.
\end{equation}
Combining the index relation \eqref{index:np} with the constraint \eqref{p:constraint}, we find the constraint on the potential power index, 
\begin{equation}
    1.02 < n < 5.87.
\end{equation}
For integer values of $n$, we find:  $n = 2$ for $p = 3/2$,
 $n = 3$ for $p = 8/5$,
  $n = 4$ for $p = 5/3$,
  and $n=5$ for $p=12/7$. 
We also examine the minimal case with $n = 1$ for $p=4/3$, which is disfavored by the P-ACT-LB data but lies near the upper bound of the allowed range.   This case may still be consistent with observations when the combined constraints on both $n_s$ and $r$ are taken into account. These joint constraints are derived from the P-ACT-LB dataset and the B-mode polarization measurements from the BICEP/Keck Array experiments at the South Pole (BK18) \cite{BICEP:2021xfz}, hereafter referred to as P-ACT-LB-BK18 \cite{ACT:2025tim}.


\subsection{Model-independent characteristic}
In deriving the KKLT potential \eqref{brane1}, we employed a reconstruction method with the condition $C > 1$ to ensure a sufficiently small tensor-to-scalar ratio compatible with current observational limits. Apart from the commonly used slow-roll condition,   no additional assumptions or model-specific inputs were introduced.   As a result, the reconstructed potential exhibits a certain degree of universality, satisfying the ACT observational constraints in a model-independent manner.
Inflationary scenarios that yield a scalar spectral index of the form $n_s = 1 - p/N$ and are simultaneously consistent with the observational bounds on both $n_s$ and $r$ are expected to give rise to potentials that, when expressed in the Einstein frame, closely resemble the reconstructed form in Eq.~\eqref{brane1}, at least on large scales.

A concrete example is provided by the inflation model with nonminimal coupling  in the Jordan frame, characterized by the coupling function $\Omega(\phi) = 1 + \xi f(\phi)$ and potential $V_J(\phi) = \lambda^2 f(\phi)$, where $f(\phi) = \phi^k$ with $k < 2$. For these models, the resulting spectral observables are given by \cite{Gao:2025onc}:
\begin{equation}\label{nsr_an}
n_s -1 \approx -\frac{k + 2}{2N}, \quad r \approx \frac{8 k^{1 - k/2}}{2^k \, \xi \, N^{k/2 + 1}},
\end{equation}
where $\xi \sim \mathcal{O}(1)$.  These predictions are consistent with those obtained from the reconstructed KKLT potential \eqref{brane1}. 
Comparing Eq. \eqref{nsr_an} with our parametrized predictions in Eq.~\eqref{nsr:para}, we find a correspondence between the two parameter sets   
\begin{equation}
    k=2p-2, \quad \xi = \frac{2^{2-p}}{[(2-p)M]^{2p-2}}.
\end{equation}
This correspondence demonstrates that the Einstein-frame potential derived from the inflation model with nonminimal coupling  coincides with the reconstructed KKLT potential in the slow-roll regime, reflecting the underlying model-independence of Eq.~\eqref{brane1}.  The Einstein-frame potential for the nonminimal model is obtained via a conformal transformation:
\begin{gather}
\label{conftransf1} 
d\psi^2 = \left[\frac{3}{2} \frac{(d\Omega/d\phi)^2}{\Omega^2(\phi)} + \frac{\omega(\phi)}{\Omega(\phi)} \right] d\phi^2, \quad V_E(\psi) =   \frac{V_J(\phi)}{\Omega^2(\phi)}.
\end{gather}

\begin{figure}[htbp]
    \centering
    \includegraphics[width=0.5\textwidth]{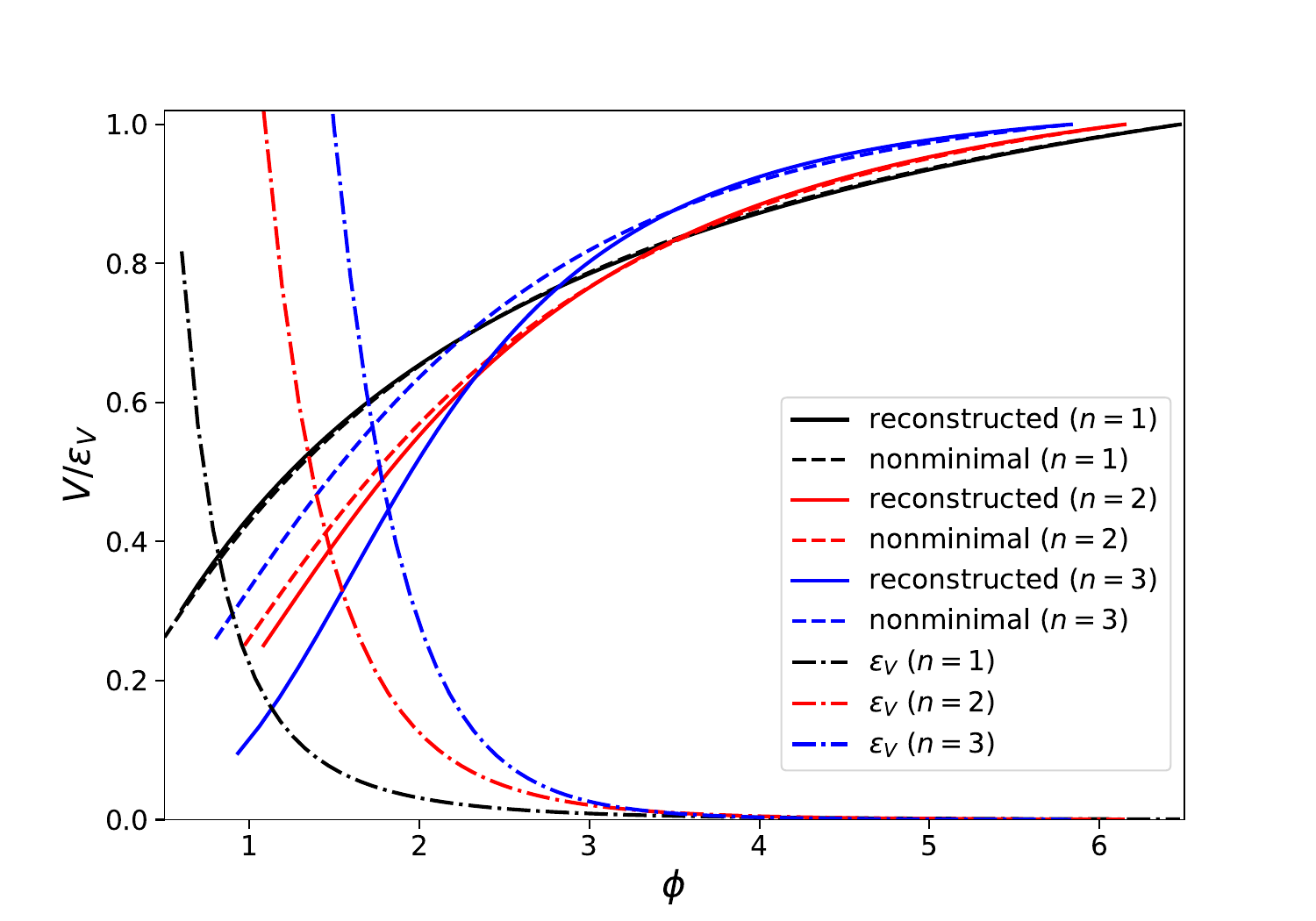}
    \caption{Comparison of the reconstructed KKLT potential \eqref{brane1} and the Einstein-frame potential of the nonminimally coupled model. Solid curves represent the reconstructed potential, while dashed curves correspond to the Einstein-frame potential of the inflation model with the nonminimal coupling. The black, red, and blue curves denote the  potentials with $n=1$, $n=2$, and $n=3$, respectively.  In all cases, we set $M = 2$. The two potentials coincide in the slow-roll region, confirming the model-independent nature of the reconstructed form, at least on large scales. For the case with $n=1$, the two potentials are nearly indistinguishable throughout the entire field range.  Dot-dashed curves show the slow-roll parameter $\epsilon_V$ of the reconstructed potential.}
    \label{fig:potential}
\end{figure}

In Fig.~\ref{fig:potential}, we compare the reconstructed KKLT potential~\eqref{brane1} (solid curves) with the Einstein-frame potential of the nonminimally coupled inflation model (dashed curves). All potentials are rescaled for clarity and ease of visualization. We present three representative cases: $n = 1$, $n = 2$, and $n = 3$, shown by the black, red, and blue curves, respectively. In all cases, we set $M = 2$. The dot-dashed curves represent the corresponding slow-roll parameter $\epsilon_V$  of the reconstructed potentials.

For $n = 1$, the reconstructed potential and the Einstein-frame potential of the nonminimally coupled model are virtually indistinguishable across the entire field range.  For $n = 2$ and $n = 3$, the agreement persists only within the slow-roll region,  while deviations arise at small field values where the slow-roll approximation begins to break down.   The prediction  \eqref{nsr_an} given by the inflation model with nonminimal coupling  holds  under the condition $k < 2$,  as it relies on an approximation that becomes increasingly accurate for smaller values of $k$. As a result, for $n = 1$, corresponding to a small $k$ with $k = 2/3$, the two potentials  become nearly indistinguishable across the entire field range. In contrast, for slightly larger values of $k$, such as $n = 2$ (i.e., $k = 1$) and $n=3$ (i.e., $k = 6/5$), the agreement between the two potentials is restricted to the slow-roll region, while significant deviations appear beyond this region.  

\subsection{Polynomial $\alpha$ attractors}

The reconstructed potential, approximated by an inverse-power plateau  form \eqref{rec:potential}, corresponds to polynomial $\alpha$-attractors  \cite{Kallosh:2022feu} or D-brane inflation \cite{Kallosh:2019jnl}. 
As shown in Eq.~\eqref{nsr:para}, the scalar spectral index $n_s$ depends solely on the power index $n$ (through the parameter $p$) and is independent of the scale parameter $M$. In contrast, the tensor-to-scalar ratio $r$ is sensitive to $M$ and follows the relation $r \propto M^{2p - 2}/N^p$. Hence, by taking sufficiently small values of $M$, the tensor-to-scalar ratio can be made arbitrarily small to be consistent with observational constraints.

\begin{figure}[htbp]
    \centering
\includegraphics[width=0.5\textwidth]{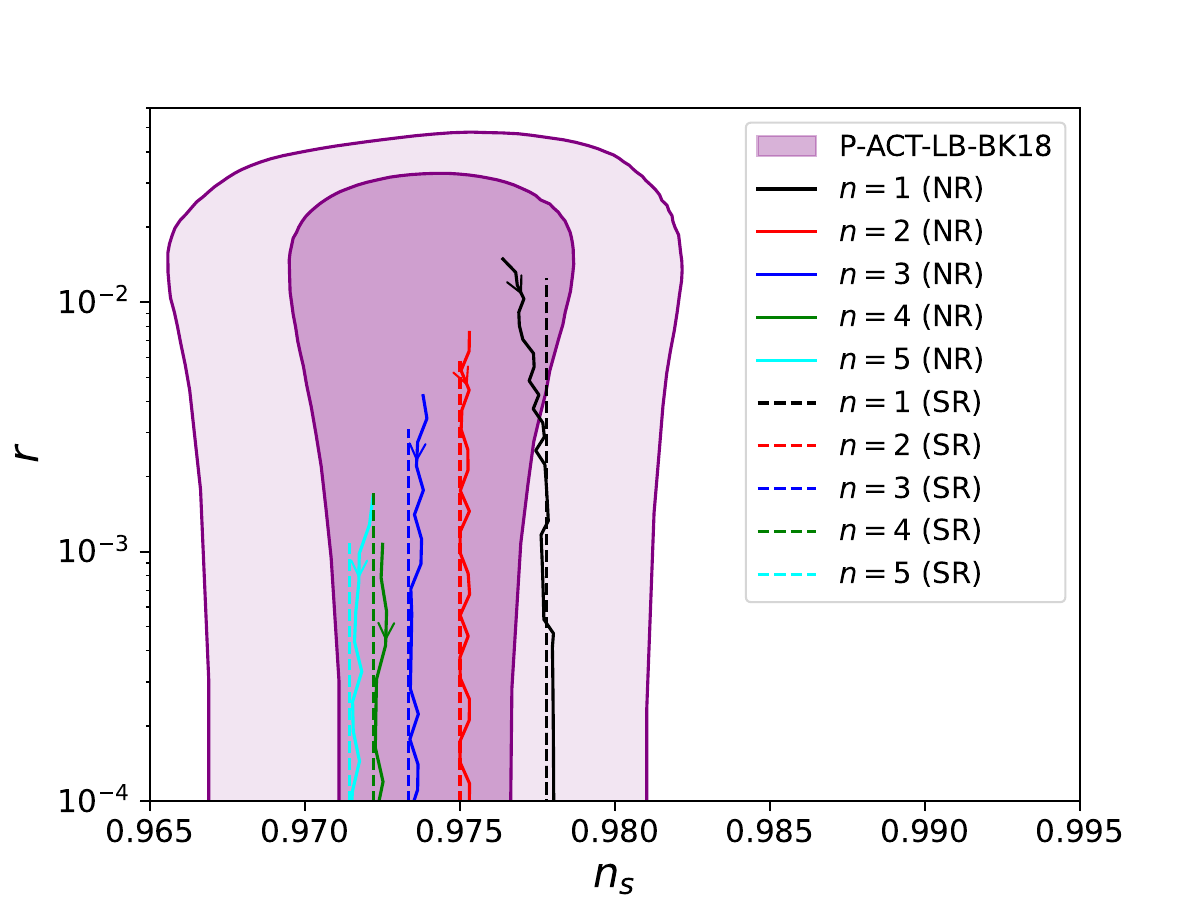}
    \caption{
    The scalar spectral index $n_s$ and tensor-to-scalar ratio $r$ as functions of the parameter $n$ for various values of $M$. Arrows indicate the direction of decreasing $M$ within the range $0.001 < M < 2$. Solid curves represent the numerical results from the polynomial $\alpha$ attractors potential \eqref{rec:potential}, while dashed curves correspond to the slow-roll approximation given by Eq.~\eqref{nsr:para}. The black, red, blue,  green, and cyan curves correspond to $n = 1$, $n = 2$, $n = 3$, $n = 4$, and $n=5$, respectively. }\label{fig:attractor}
\end{figure}

To test the predictions of the reconstructed polynomial $\alpha$-attractors  potential   \eqref{rec:potential}, we numerically solve both the background evolution and the Mukhanov-Sasaki equation to compute the scalar spectral index $n_s$ and tensor-to-scalar ratio $r$. The results are presented in  Fig.~\ref{fig:attractor}.
In Fig.~\ref{fig:attractor}, the solid black, red, blue, green, and cyan curves represent the numerical results (NR) for $n_s$ and $r$ corresponding to $n = 1$, $2$, $3$, $4$, and $5$, respectively.  The corresponding dashed curves show the analytical predictions from the slow-roll approximation (SR) given in Eq.~\eqref{nsr:para}. In all cases, we fix the number of $e$-folds to $N = 60$. Arrows along the curves indicate the direction of decreasing $M$.

The numerical results closely match the slow-roll predictions, indicating the validity of the slow-roll approximation within the considered parameter space. For $n = 2$ to $n = 5$, the results are fully consistent with the latest observational constraints. Interestingly, for the simplest case $n = 1$, it is disfavored by the P-ACT-LB observational constraints, but remains observationally viable when both $n_s$ and $r$ are simultaneously taken into account.

To realize the attractor behavior described by Eq.~\eqref{nsr:para} and the corresponding reconstructed potential \eqref{rec:potential}, we assume a large integration constant $C$,  satisfying $C > 1$. This assumption is well motivated, as the tensor-to-scalar ratio $r$ scales inversely with $C$, implying that a large $C$ naturally yields a small $r$, consistent with current observational bounds.

\subsection{The reheating}
In the above analysis we have fixed the inflationary $e$-folding number to $N=60$ without taking the reheating phase into account.  However, in more general scenarios where reheating proceeds over a finite duration with an effective equation of state, the total number of $e$-folds during inflation can vary significantly.  Allowing for this freedom effectively shifts the predicted values of $(n_s,r)$ and consequently modifies the constraints on the model parameter $p$.

Assuming that the reheating epoch follows immediately after inflation with a constant equation of state $w_{re}$, and that radiation domination begins right after reheating ends, the relation between the inflationary $e$-folding number $N$ and the reheating $e$-folding number $N_{re}$ can be written as \cite{Dai:2014jja,Cook:2015vqa}  
\begin{equation}
 \label{Nre}
    N_{re} = \frac{4}{1-3 w_{re}}\left[61.6-N_*-\ln\left(\frac{V_e^{1/4}}{H_*}\right)\right],
\end{equation}
where $V_e$ is the potential at the end of inflation and $H_*$ is the Hubble parameter at the horizon crossing of the pivot scale. Here and in what follows, the subscript “$*$” denotes quantities evaluated at horizon crossing.  The temperature at the end of reheating is given by
\begin{equation}\label{eq:reh}
    T_{re}=\exp\left[-\frac{3N_{re}(1+w_{re})}{4}\right]\left[\frac{45V_{e}}{\pi^2 g_{re}}\right]^{1/4},
\end{equation}
with the effective number of relativistic degrees of freedom $g_{re}= 106.75$.

Using Eq.~\eqref{eq:mphi} and the slow-roll background relation
\begin{equation}
    H_*^2 \approx \frac{1}{3}V(\phi_*),
\end{equation}
we obtain
\begin{equation}\label{reheating:Nre2}
    N_{re} = \frac{4}{1-3w_{re}}\left[57.4-N_* + \frac{1}{4}\ln \epsilon_{V*}-\frac{1}{4}\ln\left( \frac{1+2/[C(N_*+\alpha)^{p-1}]}{1+2/(C \alpha^{p-1})}\right) \right],
\end{equation}
and 
\begin{equation}
    T_{re} =0.012 \exp\left[-\frac{3N_{re}(1+w_{re})}{4}\right]\left[ \frac{1+2/[C(N_*+\alpha)^{p-1}]}{1+2/(C \alpha^{p-1})}\epsilon_{V*} \right]^{1/4},
\end{equation}
where we use the relation $H_*^2 = 8\pi^2 \epsilon_{V*} A_s$ with $\ln (10^{10} A_s)= 3.044$ \cite{Planck:2018jri}.  

Equation~\eqref{reheating:Nre2} shows that for a soft equation of state $w_{re}<1/3$, increasing the reheating $e$-folding number $N_{re}$ reduces the inflationary $e$-folding number $N$. Conversely, for a stiff equation of state $w_{re}>1/3$, increasing $N_{re}$ leads to a larger $N$. Since $N$ and $p$ are correlated through the observational constraint~\eqref{ns:constraint}, a larger $N$ implies a larger $p$, while a smaller $N$ implies a smaller $p$.

\begin{figure}[htbp]
    \centering
\includegraphics[width=0.45\textwidth]{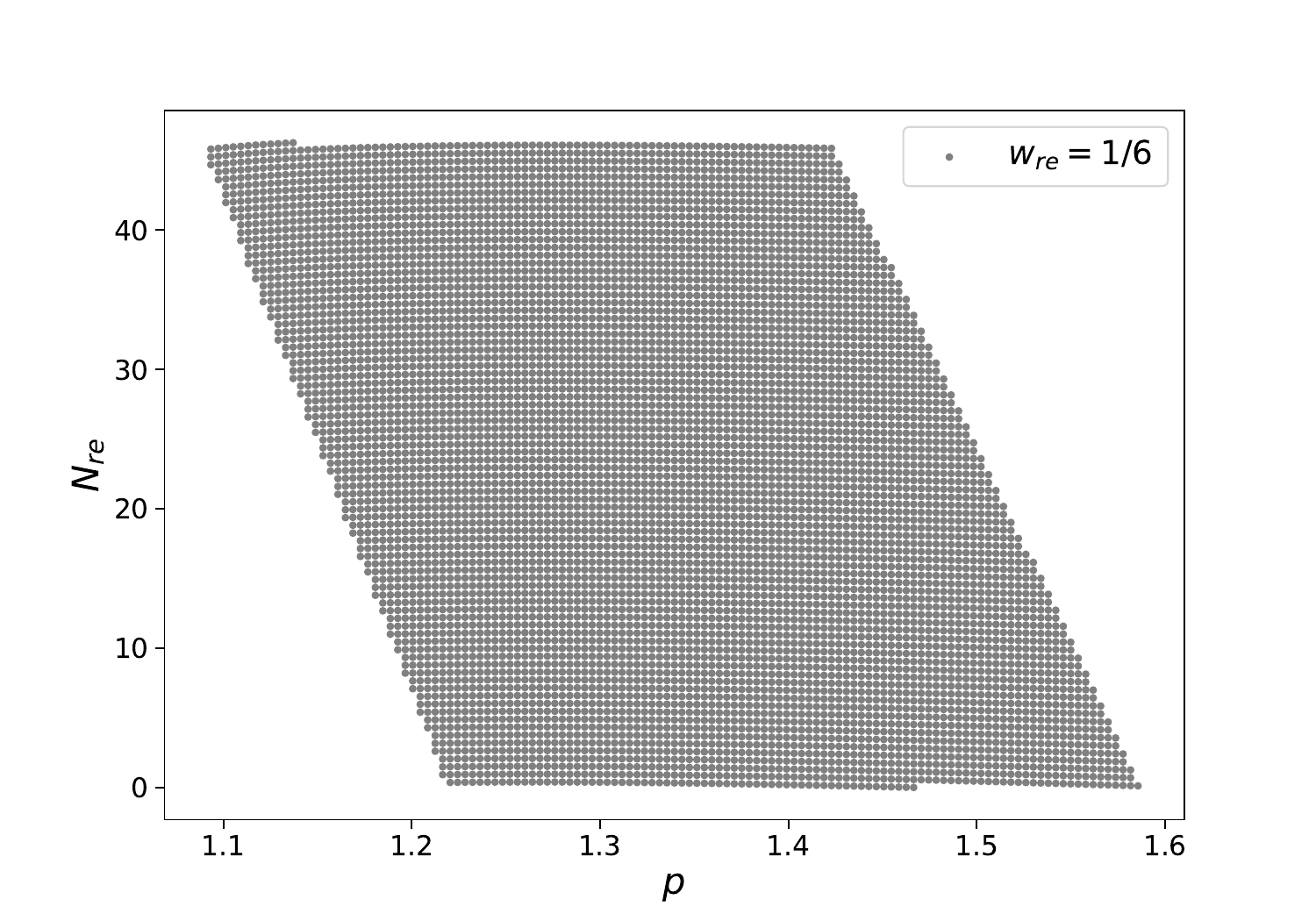}
\includegraphics[width=0.45\textwidth]{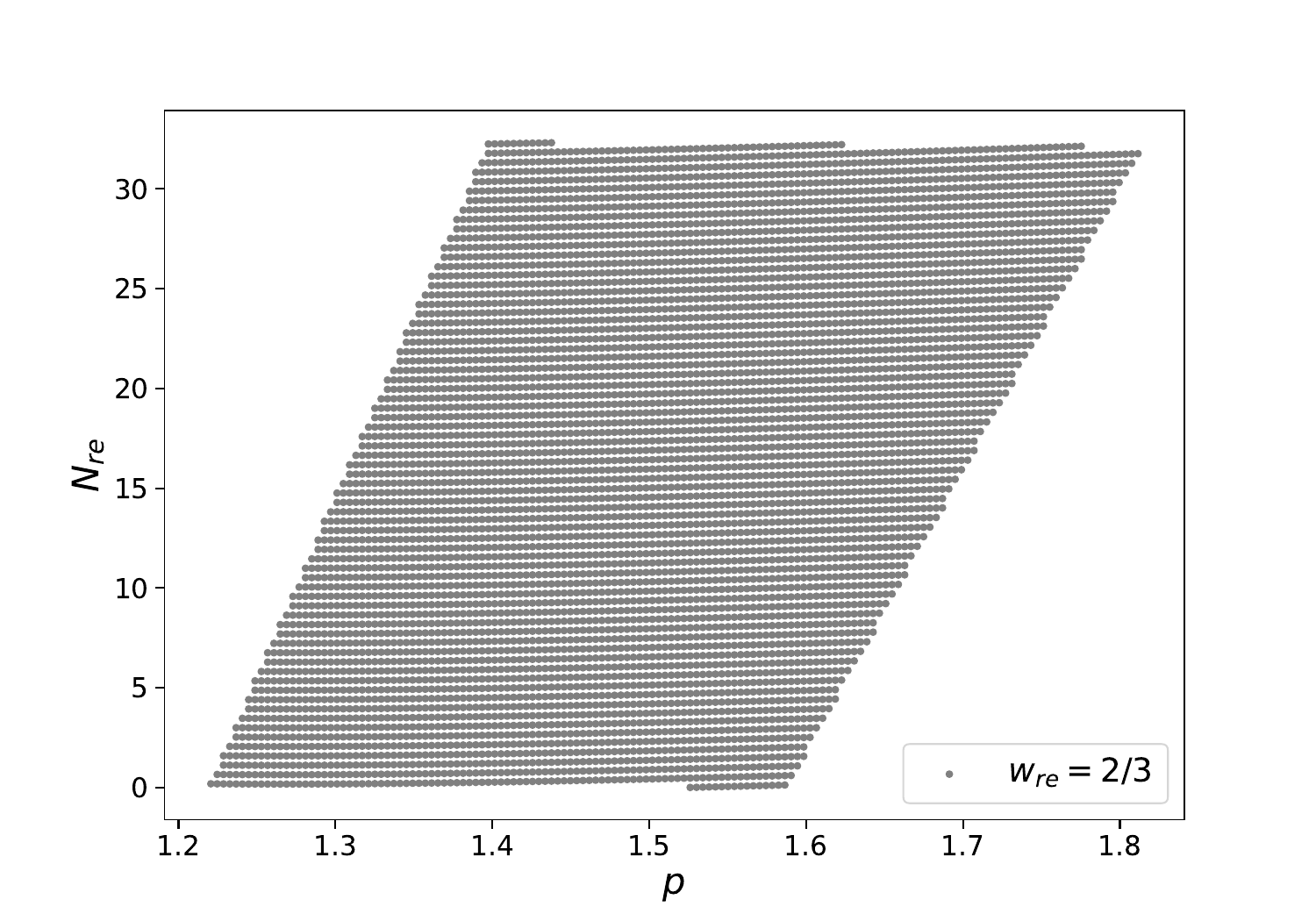}
    \caption{Constraints on the reheating $e$-folding number $N_{re}$ and the parameter $p$ from the observational constraint \eqref{ns:constraint}. 
    The left panel corresponds to a soft equation of state $w_{re}= 1/6$ and the right panel to a stiff equation of state $w_{re}=2/3$.
      }\label{fig:reheating}
\end{figure}

To quantify the impact of reheating on the constraint of $p$, we consider two representative cases: $w_{re}= 1/6$ for soft reheating and $w_{re} = 2/3$ for stiff reheating. 
The allowed regions in the $(N_{re},p)$ plane are shown in Fig.~\ref{fig:reheating}, where the left panel corresponds to $w_{re}=1/6$ and the right panel to $w_{re}=2/3$. 
In this analysis, we fix $C=100$ to ensure a sufficiently small tensor-to-scalar ratio consistent with observations, and impose the Big Bang Nucleosynthesis (BBN) bound $T_{re}>4~\mathrm{MeV}$.

For instantaneous reheating ($N_{re}=0$), the required number of inflationary $e$-folds is $N\approx 54.6$, and the observational constraint on $p$ shifts to $1.22<p<1.59$. 
Soft reheating shifts the allowed $p$ toward smaller values, while stiff reheating shifts it toward larger values. 
In particular, maintaining $N=60$ during inflation requires a reheating stage with $N_{re}\approx 22$ for $w_{re}=2/3$.

\section{Discussion and Conclusion}
\label{sec:con}
The latest observational data from the Atacama Cosmology Telescope (ACT) have provided updated constraints on the scalar spectral index $n_s$, with the P-ACT-LB results indicating a higher value of $n_s$ than previous {\it Planck} measurements. This shift poses a serious challenge to the commonly used parameterization $n_s = 1 - 2/N$, which serves as a universal attractor for a wide class of inflationary models, including hilltop inflation, the T and E models, the Starobinsky model, and Higgs inflation with nonminimal coupling. As a result, these attractor models are now disfavored by current observational data.

In this work, we have applied the potential reconstruction method to determine inflationary potentials consistent with the latest ACT data. Specifically, we consider a parameterization of the scalar spectral index of the form $n_s - 1 = -p/(N+\alpha)$. 
In the CMB scales, $N\sim 50-60$, so $\alpha\ll N$.
Taking $N = 60$ and $\alpha \approx 0$, consistency with the P-ACT-LB constraints requires the parameter $p$ to lie in the range $1.338 < p < 1.746$. The corresponding reconstructed potential is the KKLT potential $ V(\phi)=V_0/\left[1+\left(M/\phi\right)^n\right]$,  
where the power index is given by $n = 2(p-1)/(2-p)$. 
The tensor-to-scalar ratio is expressed as  $r \approx  16(p-1)/(C N^p)$ with the constant $C = 2^{2p - 1} \left[\sqrt{p - 1}/(2 - p)\right]^{2p - 2} / M^{2p - 2}$. To satisfy observational constraints, the integer values of the power index $n$ can be taken as $1$, $2$, $3$, $4$, and $5$, corresponding to $p = 4/3$, $3/2$, $8/5$,  $5/3$, and $12/7$, respectively.

Moreover, in the large $\phi$ region, the reconstructed models become the D-brane inflation and the polynomial $\alpha$-attractors. 
In these models, the scalar spectral index $n_s$ depends solely on the power index $n$ and remains invariant under changes in the parameter $M$. In contrast, the tensor-to-scalar ratio scales as $r \propto M^{2p - 2}/N^p$, and thus can be made arbitrarily small by choosing sufficiently small $M$, ensuring compatibility with the latest observational limits on $r$.

In conclusion, guided by the latest ACT observations and working within the slow-roll approximation, 
the reconstruction method reveals that the KKLT model serves as an effective approximation to a broad class of inflationary scenarios consistent with the observed value of $n_s$ on large scales. 
Moreover, this potential also approximates polynomial $\alpha$-attractors, whose attractor properties remain flexible and adaptable to future observational constraints.

\section*{Acknowledgments}
This work is supported in part by the National Key Research and Development Program of China under Grant No. 2020YFC2201504, the National Natural Science Foundation of China under Grant No. 12175184, No. 12205015, No. 12305075, the National Natural Science Foundation of China key project under Grant No. 12535002, the supporting fund for Young Researcher of Beijing Normal University under Grant No. 28719/310432102, the Chongqing Natural Science Foundation under Grant No. CSTB2022NSCQ-MSX1324 and the Startup Research Fund of Henan Academy of Science under Grant No. 241841223.

\bibliographystyle{elsarticle-num}
\bibliography{ref}

@article{Starobinsky:1980te,
    author = "Starobinsky, Alexei A.",
    editor = "Khalatnikov, I. M. and Mineev, V. P.",
    title = "{A New Type of Isotropic Cosmological Models Without Singularity}",
    doi = "10.1016/0370-2693(80)90670-X",
    journal = "Phys. Lett. B",
    volume = "91",
    pages = "99--102",
    year = "1980"
}

@article{Guth:1980zm,
    author = "Guth, Alan H.",
    editor = "Fang, Li-Zhi and Ruffini, R.",
    title = "{The Inflationary Universe: A Possible Solution to the Horizon and Flatness Problems}",
    reportNumber = "SLAC-PUB-2576",
    doi = "10.1103/PhysRevD.23.347",
    journal = "Phys. Rev. D",
    volume = "23",
    pages = "347--356",
    year = "1981"
}

@article{Linde:1981mu,
    author = "Linde, Andrei D.",
    editor = "Fang, Li-Zhi and Ruffini, R.",
    title = "{A New Inflationary Universe Scenario: A Possible Solution of the Horizon, Flatness, Homogeneity, Isotropy and Primordial Monopole Problems}",
    reportNumber = "LEBEDEV-81-229",
    doi = "10.1016/0370-2693(82)91219-9",
    journal = "Phys. Lett. B",
    volume = "108",
    pages = "389--393",
    year = "1982"
}

@article{Albrecht:1982wi,
    author = "Albrecht, Andreas and Steinhardt, Paul J.",
    editor = "Fang, Li-Zhi and Ruffini, R.",
    title = "{Cosmology for Grand Unified Theories with Radiatively Induced Symmetry Breaking}",
    reportNumber = "UPR-0185T",
    doi = "10.1103/PhysRevLett.48.1220",
    journal = "Phys. Rev. Lett.",
    volume = "48",
    pages = "1220--1223",
    year = "1982"
}

@article{ACT:2025fju,
    author = "Louis, Thibaut and others",
    collaboration = "ACT",
    title = "{The Atacama Cosmology Telescope: DR6 Power Spectra, Likelihoods and $\Lambda$CDM Parameters}",
    eprint = "2503.14452",
    archivePrefix = "arXiv",
    primaryClass = "astro-ph.CO",
    reportNumber = "FERMILAB-PUB-25-0071-PPD",
    month = "3",
    year = "2025"
}

@article{ACT:2025tim,
    author = "Calabrese, Erminia and others",
    collaboration = "ACT",
    title = "{The Atacama Cosmology Telescope: DR6 Constraints on Extended Cosmological Models}",
    eprint = "2503.14454",
    archivePrefix = "arXiv",
    primaryClass = "astro-ph.CO",
    reportNumber = "FERMILAB-PUB-25-0157-PPD",
    month = "3",
    year = "2025"
}

@article{Planck:2018jri,
    author = "Akrami, Y. and others",
    collaboration = "Planck",
    title = "{Planck 2018 results. X. Constraints on inflation}",
    eprint = "1807.06211",
    archivePrefix = "arXiv",
    primaryClass = "astro-ph.CO",
    doi = "10.1051/0004-6361/201833887",
    journal = "Astron. Astrophys.",
    volume = "641",
    pages = "A10",
    year = "2020"
}

@article{DESI:2024uvr,
    author = "Adame, A. G. and others",
    collaboration = "DESI",
    title = "{DESI 2024 III: baryon acoustic oscillations from galaxies and quasars}",
    eprint = "2404.03000",
    archivePrefix = "arXiv",
    primaryClass = "astro-ph.CO",
    reportNumber = "FERMILAB-PUB-24-0159-PPD",
    doi = "10.1088/1475-7516/2025/04/012",
    journal = "JCAP",
    volume = "04",
    pages = "012",
    year = "2025"
}

@article{DESI:2024mwx,
    author = "Adame, A. G. and others",
    collaboration = "DESI",
    title = "{DESI 2024 VI: cosmological constraints from the measurements of baryon acoustic oscillations}",
    eprint = "2404.03002",
    archivePrefix = "arXiv",
    primaryClass = "astro-ph.CO",
    reportNumber = "FERMILAB-PUB-24-0154-PPD",
    doi = "10.1088/1475-7516/2025/02/021",
    journal = "JCAP",
    volume = "02",
    pages = "021",
    year = "2025"
}

@article{Ferreira:2025lrd,
    author = "Ferreira, Elisa G. M. and McDonough, Evan and Balkenhol, Lennart and Kallosh, Renata and Knox, Lloyd and Linde, Andrei",
    title = "{The BAO-CMB Tension and Implications for Inflation}",
    eprint = "2507.12459",
    archivePrefix = "arXiv",
    primaryClass = "astro-ph.CO",
    month = "7",
    year = "2025"
}

@article{Boubekeur:2005zm,
    author = "Boubekeur, Lotfi and Lyth, David. H.",
    title = "{Hilltop inflation}",
    eprint = "hep-ph/0502047",
    archivePrefix = "arXiv",
    doi = "10.1088/1475-7516/2005/07/010",
    journal = "JCAP",
    volume = "07",
    pages = "010",
    year = "2005"
}

@article{Kallosh:2013hoa,
    author = "Kallosh, Renata and Linde, Andrei",
    title = "{Universality Class in Conformal Inflation}",
    eprint = "1306.5220",
    archivePrefix = "arXiv",
    primaryClass = "hep-th",
    doi = "10.1088/1475-7516/2013/07/002",
    journal = "JCAP",
    volume = "07",
    pages = "002",
    year = "2013"
}

@article{Kallosh:2013maa,
    author = "Kallosh, Renata and Linde, Andrei",
    title = "{Non-minimal Inflationary Attractors}",
    eprint = "1307.7938",
    archivePrefix = "arXiv",
    primaryClass = "hep-th",
    doi = "10.1088/1475-7516/2013/10/033",
    journal = "JCAP",
    volume = "10",
    pages = "033",
    year = "2013"
}

@article{Kaiser:1994vs,
    author = "Kaiser, David I.",
    title = "{Primordial spectral indices from generalized Einstein theories}",
    eprint = "astro-ph/9408044",
    archivePrefix = "arXiv",
    reportNumber = "DART-HEP-94-05, HUTP-94-A033",
    doi = "10.1103/PhysRevD.52.4295",
    journal = "Phys. Rev. D",
    volume = "52",
    pages = "4295--4306",
    year = "1995"
}

@article{Bezrukov:2007ep,
    author = "Bezrukov, Fedor L. and Shaposhnikov, Mikhail",
    title = "{The Standard Model Higgs boson as the inflaton}",
    eprint = "0710.3755",
    archivePrefix = "arXiv",
    primaryClass = "hep-th",
    doi = "10.1016/j.physletb.2007.11.072",
    journal = "Phys. Lett. B",
    volume = "659",
    pages = "703--706",
    year = "2008"
}

@article{Kallosh:2025rni,
    author = "Kallosh, Renata and Linde, Andrei and Roest, Diederik",
    title = "{Atacama Cosmology Telescope, South Pole Telescope, and Chaotic Inflation}",
    eprint = "2503.21030",
    archivePrefix = "arXiv",
    primaryClass = "hep-th",
    doi = "10.1103/d6gn-78hn",
    journal = "Phys. Rev. Lett.",
    volume = "135",
    number = "16",
    pages = "161001",
    year = "2025"
}

@article{Gao:2025onc,
    author = "Gao, Qing and Gong, Yungui and Yi, Zhu and Zhang, Fengge",
    title = "{Nonminimal coupling in light of ACT data}",
    eprint = "2504.15218",
    archivePrefix = "arXiv",
    primaryClass = "astro-ph.CO",
    doi = "10.1016/j.dark.2025.102106",
    journal = "Phys. Dark Univ.",
    volume = "50",
    pages = "102106",
    year = "2025"
}

@article{Liu:2025qca,
    author = "Liu, Lang and Yi, Zhu and Gong, Yungui",
    title = "{Reconciling Higgs Inflation with ACT Observations through Reheating}",
    eprint = "2505.02407",
    archivePrefix = "arXiv",
    primaryClass = "astro-ph.CO",
    month = "5",
    year = "2025"
}

@article{Haque:2025uis,
    author = "Haque, Md Riajul and Pal, Sourav and Paul, Debarun",
    title = "{Improved predictions on Higgs-Starobinsky inflation and reheating with ACT DR6 and primordial gravitational waves}",
    eprint = "2505.04615",
    archivePrefix = "arXiv",
    primaryClass = "astro-ph.CO",
    doi = "10.1016/j.physletb.2025.139852",
    journal = "Phys. Lett. B",
    volume = "869",
    pages = "139852",
    year = "2025"
}

@article{Zharov:2025evb,
    author = "Zharov, D. S. and Sobol, O. O. and Vilchinskii, S. I.",
    title = "{Reheating ACTs on Starobinsky and Higgs inflation}",
    eprint = "2505.01129",
    archivePrefix = "arXiv",
    primaryClass = "astro-ph.CO",
    month = "5",
    year = "2025"
}

@article{Haque:2025uri,
    author = "Haque, Md Riajul and Pal, Sourav and Paul, Debarun",
    title = "{ACT DR6 Insights on the Inflationary Attractor models and Reheating}",
    eprint = "2505.01517",
    archivePrefix = "arXiv",
    primaryClass = "astro-ph.CO",
    month = "5",
    year = "2025"
}

@article{Drees:2025ngb,
    author = "Drees, Manuel and Xu, Yong",
    title = "{Refined predictions for Starobinsky inflation and post-inflationary constraints in light of ACT}",
    eprint = "2504.20757",
    archivePrefix = "arXiv",
    primaryClass = "astro-ph.CO",
    reportNumber = "MITP-25-033",
    doi = "10.1016/j.physletb.2025.139612",
    journal = "Phys. Lett. B",
    volume = "867",
    pages = "139612",
    year = "2025"
}

@article{Ballardini:2024ado,
    author = "Ballardini, Mario",
    title = "{Chasing cosmic inflation: constraints for inflationary models and reheating insights}",
    eprint = "2408.03321",
    archivePrefix = "arXiv",
    primaryClass = "astro-ph.CO",
    doi = "10.1088/1475-7516/2025/01/116",
    journal = "JCAP",
    volume = "01",
    pages = "116",
    year = "2025"
}

@article{He:2025bli,
    author = "He, Minxi and Hong, Muzi and Mukaida, Kyohei",
    title = "{Increase of ns in regularized pole inflation {\&} Einstein-Cartan gravity}",
    eprint = "2504.16069",
    archivePrefix = "arXiv",
    primaryClass = "astro-ph.CO",
    doi = "10.1088/1475-7516/2025/09/080",
    journal = "JCAP",
    volume = "09",
    pages = "080",
    year = "2025"
}

@article{Gialamas:2025kef,
    author = "Gialamas, Ioannis D. and Karam, Alexandros and Racioppi, Antonio and Raidal, Martti",
    title = "{Has ACT measured radiative corrections to the tree-level Higgs-like inflation?}",
    eprint = "2504.06002",
    archivePrefix = "arXiv",
    primaryClass = "astro-ph.CO",
    month = "4",
    year = "2025"
}

@article{Frob:2025sfq,
    author = {Fr{\"o}b, Markus B. and Glavan, Dra{\v{z}}en and Meda, Paolo and Sawicki, Ignacy},
    title = "{One-loop correction to primordial tensor modes during radiation era}",
    eprint = "2504.02609",
    archivePrefix = "arXiv",
    primaryClass = "astro-ph.CO",
    month = "4",
    year = "2025"
}

@article{Dioguardi:2025vci,
    author = "Dioguardi, Christian and Iovino, Antonio J. and Racioppi, Antonio",
    title = "{Fractional attractors in light of the latest ACT observations}",
    eprint = "2504.02809",
    archivePrefix = "arXiv",
    primaryClass = "gr-qc",
    doi = "10.1016/j.physletb.2025.139664",
    journal = "Phys. Lett. B",
    volume = "868",
    pages = "139664",
    year = "2025"
}

@article{Brahma:2025dio,
    author = "Brahma, Suddhasattwa and Calder{\'o}n-Figueroa, Jaime",
    title = "{Is the CMB revealing signs of pre-inflationary physics?}",
    eprint = "2504.02746",
    archivePrefix = "arXiv",
    primaryClass = "astro-ph.CO",
    month = "4",
    year = "2025"
}

@article{Berera:2025vsu,
    author = "Berera, Arjun and Brahma, Suddhasattwa and Qiu, Zizang and O. Ramos, Rudnei and Rodrigues, Gabriel S.",
    title = "{The early universe is $\textit{ACT}$-ing $\textit{warm}$}",
    eprint = "2504.02655",
    archivePrefix = "arXiv",
    primaryClass = "hep-th",
    month = "4",
    year = "2025"
}

@article{Aoki:2025wld,
    author = "Aoki, Shuntaro and Otsuka, Hajime and Yanagita, Ryota",
    title = "{Higgs-modular inflation}",
    eprint = "2504.01622",
    archivePrefix = "arXiv",
    primaryClass = "hep-ph",
    reportNumber = "RIKEN-iTHEMS-Report-25, KYUSHU-HET-317",
    doi = "10.1103/v4z9-676d",
    journal = "Phys. Rev. D",
    volume = "112",
    number = "4",
    pages = "043505",
    year = "2025"
}

@article{Dioguardi:2025mpp,
    author = "Dioguardi, Christian and Karam, Alexandros",
    title = "{Palatini linear attractors are back in action}",
    eprint = "2504.12937",
    archivePrefix = "arXiv",
    primaryClass = "gr-qc",
    doi = "10.1103/23b3-9d7q",
    journal = "Phys. Rev. D",
    volume = "111",
    number = "12",
    pages = "123521",
    year = "2025"
}

@article{Salvio:2025izr,
    author = "Salvio, Alberto",
    title = "{Independent connection in action during inflation}",
    eprint = "2504.10488",
    archivePrefix = "arXiv",
    primaryClass = "hep-ph",
    doi = "10.1103/tq3v-vy3y",
    journal = "Phys. Rev. D",
    volume = "112",
    number = "6",
    pages = "L061301",
    year = "2025"
}

@article{Yogesh:2025wak,
    author = "Yogesh and Mohammadi, Abolhassan and Wu, Qiang and Zhu, Tao",
    title = "{Starobinsky like inflation and EGB Gravity in the light of ACT}",
    eprint = "2505.05363",
    archivePrefix = "arXiv",
    primaryClass = "astro-ph.CO",
    doi = "10.1088/1475-7516/2025/10/010",
    journal = "JCAP",
    volume = "10",
    pages = "010",
    year = "2025"
}

@article{Gialamas:2025ofz,
    author = "Gialamas, Ioannis D. and Katsoulas, Theodoros and Tamvakis, Kyriakos",
    title = "{Keeping the relation between the Starobinsky model and no-scale supergravity ACTive}",
    eprint = "2505.03608",
    archivePrefix = "arXiv",
    primaryClass = "gr-qc",
    doi = "10.1088/1475-7516/2025/09/060",
    journal = "JCAP",
    volume = "09",
    pages = "060",
    year = "2025"
}

@article{Gao:2025viy,
    author = "Gao, Qing and Qian, Yanjiang and Gong, Yungui and Yi, Zhu",
    title = "{Observational constraints on inflationary models with non-minimally derivative coupling by ACT}",
    eprint = "2506.18456",
    archivePrefix = "arXiv",
    primaryClass = "gr-qc",
    doi = "10.1088/1475-7516/2025/08/083",
    journal = "JCAP",
    volume = "08",
    pages = "083",
    year = "2025"
}

@article{Peng:2025bws,
    author = "Peng, Zhi-Zhang and Chen, Zu-Cheng and Liu, Lang",
    title = "{The polynomial potential inflation in light of ACT observations}",
    eprint = "2505.12816",
    archivePrefix = "arXiv",
    primaryClass = "astro-ph.CO",
    month = "5",
    year = "2025"
}

@article{Odintsov:2025jky,
    author = "Odintsov, S. D. and Oikonomou, V. K.",
    title = "{Confronting rainbow-deformed f(R) gravity with the ACT data}",
    eprint = "2508.17358",
    archivePrefix = "arXiv",
    primaryClass = "gr-qc",
    doi = "10.1016/j.physletb.2025.139909",
    journal = "Phys. Lett. B",
    volume = "870",
    pages = "139909",
    year = "2025"
}

@article{Oikonomou:2025htz,
    author = "Oikonomou, V. K.",
    title = "{Strong gravity effects on R2-corrected single scalar field inflation and compatibility with the ACT data}",
    eprint = "2508.17363",
    archivePrefix = "arXiv",
    primaryClass = "gr-qc",
    doi = "10.1016/j.physletb.2025.139972",
    journal = "Phys. Lett. B",
    volume = "871",
    pages = "139972",
    year = "2025"
}

@article{Odintsov:2025eiv,
    author = "Odintsov, S. D. and Oikonomou, V. K.",
    title = "{Power-law F(R) gravity as deformations to Starobinsky inflation in view of ACT}",
    eprint = "2509.06251",
    archivePrefix = "arXiv",
    primaryClass = "gr-qc",
    doi = "10.1016/j.physletb.2025.139907",
    journal = "Phys. Lett. B",
    volume = "870",
    pages = "139907",
    year = "2025"
}

@article{Zhu:2025twm,
    author = "Zhu, Yigan and Gao, Qing and Gong, Yungui and Yi, Zhu",
    title = "{Inflationary models with Gauss{\textendash}Bonnet coupling in light of ACT observations}",
    eprint = "2508.09707",
    archivePrefix = "arXiv",
    primaryClass = "astro-ph.CO",
    doi = "10.1140/epjc/s10052-025-14969-2",
    journal = "Eur. Phys. J. C",
    volume = "85",
    number = "10",
    pages = "1227",
    year = "2025"
}

@article{Mukhanov:2013tua,
    author = "Mukhanov, Viatcheslav",
    title = "{Quantum Cosmological Perturbations: Predictions and Observations}",
    eprint = "1303.3925",
    archivePrefix = "arXiv",
    primaryClass = "astro-ph.CO",
    doi = "10.1140/epjc/s10052-013-2486-7",
    journal = "Eur. Phys. J. C",
    volume = "73",
    pages = "2486",
    year = "2013"
}

@article{Roest:2013fha,
    author = "Roest, Diederik",
    title = "{Universality classes of inflation}",
    eprint = "1309.1285",
    archivePrefix = "arXiv",
    primaryClass = "hep-th",
    doi = "10.1088/1475-7516/2014/01/007",
    journal = "JCAP",
    volume = "01",
    pages = "007",
    year = "2014"
}

@article{Garcia-Bellido:2014gna,
    author = "Garcia-Bellido, Juan and Roest, Diederik",
    title = "{Large-$N$ running of the spectral index of inflation}",
    eprint = "1402.2059",
    archivePrefix = "arXiv",
    primaryClass = "astro-ph.CO",
    doi = "10.1103/PhysRevD.89.103527",
    journal = "Phys. Rev. D",
    volume = "89",
    number = "10",
    pages = "103527",
    year = "2014"
}

@article{Creminelli:2014nqa,
    author = "Creminelli, Paolo and Dubovsky, Sergei and L{\'o}pez Nacir, Diana and Simonovi{\'c}, Marko and Trevisan, Gabriele and Villadoro, Giovanni and Zaldarriaga, Matias",
    title = "{Implications of the scalar tilt for the tensor-to-scalar ratio}",
    eprint = "1412.0678",
    archivePrefix = "arXiv",
    primaryClass = "astro-ph.CO",
    doi = "10.1103/PhysRevD.92.123528",
    journal = "Phys. Rev. D",
    volume = "92",
    number = "12",
    pages = "123528",
    year = "2015"
}

@article{Barranco:2014ira,
    author = "Barranco, Laura and Boubekeur, Lotfi and Mena, Olga",
    title = "{A model-independent fit to Planck and BICEP2 data}",
    eprint = "1405.7188",
    archivePrefix = "arXiv",
    primaryClass = "astro-ph.CO",
    reportNumber = "IFIC-14-37",
    doi = "10.1103/PhysRevD.90.063007",
    journal = "Phys. Rev. D",
    volume = "90",
    number = "6",
    pages = "063007",
    year = "2014"
}

@article{Galante:2014ifa,
    author = "Galante, Mario and Kallosh, Renata and Linde, Andrei and Roest, Diederik",
    title = "{Unity of Cosmological Inflation Attractors}",
    eprint = "1412.3797",
    archivePrefix = "arXiv",
    primaryClass = "hep-th",
    doi = "10.1103/PhysRevLett.114.141302",
    journal = "Phys. Rev. Lett.",
    volume = "114",
    number = "14",
    pages = "141302",
    year = "2015"
}

@article{Chiba:2015zpa,
    author = "Chiba, Takeshi",
    title = "{Reconstructing the inflaton potential from the spectral index}",
    eprint = "1504.07692",
    archivePrefix = "arXiv",
    primaryClass = "astro-ph.CO",
    doi = "10.1093/ptep/ptv090",
    journal = "PTEP",
    volume = "2015",
    number = "7",
    pages = "073E02",
    year = "2015"
}

@article{Lin:2015fqa,
    author = "Lin, Jianmang and Gao, Qing and Gong, Yungui",
    title = "{The reconstruction of inflationary potentials}",
    eprint = "1508.07145",
    archivePrefix = "arXiv",
    primaryClass = "gr-qc",
    doi = "10.1093/mnras/stw915",
    journal = "Mon. Not. Roy. Astron. Soc.",
    volume = "459",
    number = "4",
    pages = "4029--4037",
    year = "2016"
}

@article{Yi:2016jqr,
    author = "Yi, Zhu and Gong, Yungui",
    title = "{Nonminimal coupling and inflationary attractors}",
    eprint = "1608.05922",
    archivePrefix = "arXiv",
    primaryClass = "gr-qc",
    doi = "10.1103/PhysRevD.94.103527",
    journal = "Phys. Rev. D",
    volume = "94",
    number = "10",
    pages = "103527",
    year = "2016"
}

@article{Odintsov:2016vzz,
    author = "Odintsov, S. D. and Oikonomou, V. K.",
    title = "{Inflationary $\alpha$-attractors from $F(R)$ gravity}",
    eprint = "1612.01126",
    archivePrefix = "arXiv",
    primaryClass = "gr-qc",
    doi = "10.1103/PhysRevD.94.124026",
    journal = "Phys. Rev. D",
    volume = "94",
    number = "12",
    pages = "124026",
    year = "2016"
}

@article{Odintsov:2017qpp,
    author = "Odintsov, S. D. and Oikonomou, V. K.",
    title = "{Inflation with a Smooth Constant-Roll to Constant-Roll Era Transition}",
    eprint = "1704.02931",
    archivePrefix = "arXiv",
    primaryClass = "gr-qc",
    doi = "10.1103/PhysRevD.96.024029",
    journal = "Phys. Rev. D",
    volume = "96",
    number = "2",
    pages = "024029",
    year = "2017"
}

@article{Nojiri:2017qvx,
    author = "Nojiri, S. and Odintsov, S. D. and Oikonomou, V. K.",
    title = "{Constant-roll Inflation in $F(R)$ Gravity}",
    eprint = "1704.05945",
    archivePrefix = "arXiv",
    primaryClass = "gr-qc",
    doi = "10.1088/1361-6382/aa92a4",
    journal = "Class. Quant. Grav.",
    volume = "34",
    number = "24",
    pages = "245012",
    year = "2017"
}

@article{Gao:2017uja,
    author = "Gao, Qing and Gong, Yungui",
    title = "{Reconstruction of extended inflationary potentials for attractors}",
    eprint = "1703.02220",
    archivePrefix = "arXiv",
    primaryClass = "gr-qc",
    doi = "10.1140/epjp/i2018-12324-3",
    journal = "Eur. Phys. J. Plus",
    volume = "133",
    number = "11",
    pages = "491",
    year = "2018"
}

@article{Gao:2017owg,
    author = "Gao, Qing",
    title = "{Reconstruction of constant slow-roll inflation}",
    eprint = "1704.08559",
    archivePrefix = "arXiv",
    primaryClass = "astro-ph.CO",
    doi = "10.1007/s11433-017-9065-4",
    journal = "Sci. China Phys. Mech. Astron.",
    volume = "60",
    number = "9",
    pages = "090411",
    year = "2017"
}

@article{Fei:2017fub,
    author = "Fei, Qin and Gong, Yungui and Lin, Jiong and Yi, Zhu",
    title = "{The reconstruction of tachyon inflationary potentials}",
    eprint = "1705.02545",
    archivePrefix = "arXiv",
    primaryClass = "gr-qc",
    doi = "10.1088/1475-7516/2017/08/018",
    journal = "JCAP",
    volume = "08",
    pages = "018",
    year = "2017"
}

@article{Fei:2020jab,
    author = "Fei, Qin and Yi, Zhu and Yang, Yingjie",
    title = "{The Reconstruction of Non-Minimal Derivative Coupling Inflationary Potentials}",
    eprint = "2009.14819",
    archivePrefix = "arXiv",
    primaryClass = "gr-qc",
    doi = "10.3390/universe6110213",
    journal = "Universe",
    volume = "6",
    number = "11",
    pages = "213",
    year = "2020"
}

@article{BICEP:2021xfz,
    author = "Ade, P. A. R. and others",
    collaboration = "BICEP, Keck",
    title = "{Improved Constraints on Primordial Gravitational Waves using Planck, WMAP, and BICEP/Keck Observations through the 2018 Observing Season}",
    eprint = "2110.00483",
    archivePrefix = "arXiv",
    primaryClass = "astro-ph.CO",
    doi = "10.1103/PhysRevLett.127.151301",
    journal = "Phys. Rev. Lett.",
    volume = "127",
    number = "15",
    pages = "151301",
    year = "2021"
}

@article{Kachru:2003aw,
    author = "Kachru, Shamit and Kallosh, Renata and Linde, Andrei D. and Trivedi, Sandip P.",
    title = "{De Sitter vacua in string theory}",
    eprint = "hep-th/0301240",
    archivePrefix = "arXiv",
    reportNumber = "SLAC-PUB-9630, SU-ITP-03-01, TIFR-TH-03-03",
    doi = "10.1103/PhysRevD.68.046005",
    journal = "Phys. Rev. D",
    volume = "68",
    pages = "046005",
    year = "2003"
}

@article{Martin:2013tda,
    author = "Martin, Jerome and Ringeval, Christophe and Vennin, Vincent",
    title = "{Encyclop{\ae}dia Inflationaris}: {Opiparous Edition}",
    eprint = "1303.3787",
    archivePrefix = "arXiv",
    primaryClass = "astro-ph.CO",
    doi = "10.1016/j.dark.2024.101653",
    journal = "Phys. Dark Univ.",
    volume = "5-6",
    pages = "75--235",
    year = "2014"
}

@article{Dioguardi:2023jwa,
    author = "Dioguardi, Christian and Racioppi, Antonio",
    title = "{Palatini F(R,X): A new framework for inflationary attractors}",
    eprint = "2307.02963",
    archivePrefix = "arXiv",
    primaryClass = "gr-qc",
    doi = "10.1016/j.dark.2024.101509",
    journal = "Phys. Dark Univ.",
    volume = "45",
    pages = "101509",
    year = "2024"
}

@inproceedings{Dvali:2001fw,
    author = "Dvali, G. R. and Shafi, Q. and Solganik, S.",
    title = "{D-brane inflation}",
    booktitle = "{4th European Meeting From the Planck Scale to the Electroweak Scale}",
    eprint = "hep-th/0105203",
    archivePrefix = "arXiv",
    month = "5",
    year = "2001"
}

@article{Kallosh:2019jnl,
    author = "Kallosh, Renata and Linde, Andrei",
    title = "{On hilltop and brane inflation after Planck}",
    eprint = "1906.02156",
    archivePrefix = "arXiv",
    primaryClass = "hep-th",
    doi = "10.1088/1475-7516/2019/09/030",
    journal = "JCAP",
    volume = "09",
    pages = "030",
    year = "2019"
}

@article{Kallosh:2022feu,
    author = "Kallosh, Renata and Linde, Andrei",
    title = "{Polynomial {\ensuremath{\alpha}}-attractors}",
    eprint = "2202.06492",
    archivePrefix = "arXiv",
    primaryClass = "astro-ph.CO",
    doi = "10.1088/1475-7516/2022/04/017",
    journal = "JCAP",
    volume = "04",
    number = "04",
    pages = "017",
    year = "2022"
}

@article{Dai:2014jja,
    author = "Dai, Liang and Kamionkowski, Marc and Wang, Junpu",
    title = "{Reheating constraints to inflationary models}",
    eprint = "1404.6704",
    archivePrefix = "arXiv",
    primaryClass = "astro-ph.CO",
    doi = "10.1103/PhysRevLett.113.041302",
    journal = "Phys. Rev. Lett.",
    volume = "113",
    pages = "041302",
    year = "2014"
}

@article{Cook:2015vqa,
    author = "Cook, Jessica L. and Dimastrogiovanni, Emanuela and Easson, Damien A. and Krauss, Lawrence M.",
    title = "{Reheating predictions in single field inflation}",
    eprint = "1502.04673",
    archivePrefix = "arXiv",
    primaryClass = "astro-ph.CO",
    doi = "10.1088/1475-7516/2015/04/047",
    journal = "JCAP",
    volume = "04",
    pages = "047",
    year = "2015"
}

\end{document}